\documentclass[a4paper]{article}

\usepackage{icrc2013}
\usepackage{rotating}
\title{Dark Matter Annihilation Limits from Dwarf Galaxies using VERITAS}

\shorttitle{ICRC 2013 VERITAS DM Limits}

\authors{
Benjamin Zitzer$^{1}$,
for the VERITAS Collaboration.
}

\afiliations{
$^1$ Argonne National Laboratory, 9700 S. Cass Ave. Lemont, IL, USA \\

}

\email{bzitzer@anl.gov}

\abstract{Current cosmological models and data suggest the existence of a cold Dark Matter (DM) component, however the nature of DM particles remains unknown. A favored candidate for DM is a Weakly Interacting Massive Particle (WIMP) in the mass range of 50 GeV to greater than 10 TeV. Nearby dwarf spheroidal galaxies (dSphs) are expected to contain a high density of DM with a low gamma-ray background, and are thus promising targets for the detection of secondary gamma rays at very high energies (VHE, $E> 0.1 $ TeV) through the annihilation of WIMPs into Standard Model (SM) particles. The VERITAS array of Imaging Atmospheric Cherenkov Telescopes (IACTs), sensitive to gamma rays in the 100GeV to 50 TeV range, carries out an extensive observation program of dSphs. Presented here are results of the observations and new statistical techniques for constraining properties of WIMP DM models.
}


\begin{document}
\maketitle

\section{Introduction}

The search for Standard Model particles resulting from the annihilation of Dark Matter particles provides an important complement to that of direct searches for DM interactions and accelerator production experiments. Among the theoretical candidates for the DM particle \cite{bib:Bertone05}; weakly interacting massive particles are well motivated since they naturally provide the measured present day cold DM density \cite{bib:Komatsu11}. Candidates for WIMP dark matter are present in many extensions of the SM of particle physics, such as supersymmetry (SUSY) \cite{bib:Jungman96} or theories with extra dimensions \cite{bib:Servant03}. In such models, the WIMPs either decay or self-annihilate into standard model particles, most of which produce either a continuum of $\gamma$-rays with energies up to the DM particle mass, or mono-energetic $\gamma$-ray lines.

Attractive targets for indirect DM searches are nearby massive objects with high inferred DM density, which are not expected to be sources of VHE $\gamma$-rays. The Galactic Center is likely the brightest source of $\gamma$-rays resulting from DM annihilations, however the detected VHE $\gamma$-ray emission is coincident with the supermassive black hole Sgr A* and a nearby pulsar wind nebula \cite{bib:Aharonian06}, motivating searches for DM annihilation in the Galactic Center halo where the VHE $\gamma$-ray background is expected to be significantly lower \cite{bib:Abramowski11}. Dwarf spheroidal galaxies (dSphs) are additional attractive targets for DM searches. Dwarf spheroidal galaxies are relatively close ($\sim$50 kpc), and have a low rate of active or recent star formation, which suggests a low background from conventional astrophysical VHE processes \cite{bib:Mateo98}. Observations of five dSphs with VERITAS are discussed here, followed by limits of the thermally-averaged neutralino self-annihilation cross section using conventional statistical methods, followed by a summary of a method of a joint DM analysis for IACTs, i.e., combining data from several dSphs into a single combined limit.

\section{Observations}

VERITAS (Very Energetic Radiation Imaging Telescope Array System) is an array of four imaging atmospheric Cherenkov telescopes (IACTs), each 12m in diameter, located at the Fred Lawrence Whipple Observatory in southern Arizona, USA. Each VERITAS camera contains 499 pixels (0.15$^{\circ}$ diameter) and has a field of view of 3.5$^{\circ}$. In the summer of 2009 the first telescope was moved to its current location in the array to provide a more uniform distance between telescopes, improving the sensitivity of the system \cite{bib:Perkins09}. VERITAS is sensitive over an energy range of 100 GeV to 30 TeV with an energy resolution of 15\%-25\% and an angular resolution (68\% containment) of less than 0.1$^{\circ}$ per event. A source with a 1\% Crab Nebula flux can be detected by VERITAS in approximately 25 hours.

Since the start of four telescope operations in 2007, five dSphs in the northern hemisphere have been observed by VERITAS: Segue 1, Ursa Minor, Draco, Bootes and Wilman 1. Quality data for this analysis requires clear, moonless, atmospheric conditions (based on infrared temperature measurements), and nominal hardware operation.  Data reduction utilized the standard methods \cite{bib:Acciari08}. Flux upper limits were calculated for each dSph since none showed the significant excess required for a detection. Results summarizing observations and preliminary analysis results are listed in Table 1.

\begin{table*}[t]
\begin{center}

\begin{tabular}[width=1.0\textwidth]{|l|p{1.5cm}|p{1.5cm}|p{1.7cm}|p{1.4cm}|p{1.9cm}|}
\hline Dwarf & Distance (kpc) & Exposure (hrs) & Significance ($\sigma$, prelim.) & $E_{th}$ (GeV, prelim.) & $F(E>E_{th})$ (CU, prelim.)  \\ \hline
Segue 1   & 23  & 83 & -1.34 & 150 & 0.15\% \\ \hline
Draco   & 80  & 38 & 0.71 & 380 & 1.36\% \\ \hline
Ursa Minor & 66  & 39 &  -1.1 & 290 & 0.52\% \\ \hline
Wilman 1 & 38 & 14 & -0.15 & 200 & 1.62\% \\ \hline
Bootes & 62 & 14 & -0.31 & 200 & 0.81\% \\ \hline
\end{tabular}
\caption{Summary of all observations and analysis of dSphs before the VERITAS camera upgrade. Analysis results are preliminary. The flux upper limit at the 95\% confidence level, in units relative to the Crab Nebula flux.}
\
\end{center}

\end{table*} 

\section{Limits on the Dark Matter Cross-Section}

\subsection{Classic Method}

The `classic' method for determining the limits of the thermally-averaged neutralino cross-section \cite{bib:Wood08} is found from counting events in the search signal region, $N_{ON}$, and one or more background regions, $N_{OFF}$, then calculating $N^{95\%CL}_{\gamma}$, the upper limit of $\gamma$-rays in the ON region.  Assuming $\gamma$-ray flux is only from DM annihilation, the limit of the velocity-averaged cross-section for a WIMP mass of $m_{\chi}$ is:

\begin{equation}
<\sigma v>^{95\%CL}=\frac{8\pi}{J(\Delta\Omega)} \frac{N^{95\%CL}_{\gamma} m^{2}_{\chi} }{t_{obs}\int^{m_{\chi}}_{0}A_{eff}(E)\frac{dN_{\gamma}}{dE}dE},
\end{equation} 
where $A_{eff}$ is the effective area function of the detector, $t_{obs}$ is the dead-time corrected exposure time of the dsph, and $\frac{dN_{\gamma}}{dE}$ is the spectrum of a single WIMP annihilation (or decay) into $\gamma$-rays. This is obtained through a monte carlo particle physics simulation for a particular annihilation channel. The integral in the denominator of equation 1 is taken from the limits 0 to the neutralino mass, $m_{\chi}$, however the effective area is zero at energies below $\sim$50 GeV. The `astrophysical factor', $J(\Delta\Omega)$, is the squared DM density (or density in the case of DM decay) integrated along the line of sight and over a solid angle defined by the VERITAS ON region. Obtaining the DM density requires a modeling of the dSph DM profile.  

This method has been successfully implemented for other works where a thermally-averaged neutralino cross-section limits are calculated \cite{bib:Wood08,bib:MAGICCollab11,bib:VERITASCollab10,bib:HESSCollab11}, including the result for 48 hour exposure on Segue 1 shown in Figure 1 \cite{bib:VERITASCollab12}. 

This method is limited because it does not use all the available photon information, i.e. the individual photon's energy or position is not used, and treats all of the events which pass cuts with equal weight in the analysis. Essentially, this is throwing all data into a single bin. The methodology described in the next section provides a more sensitive limit by using individual photon information and combining data from several dSphs into a single limit.

\begin{figure}[h]
	\centering
	\includegraphics[width=0.5\textwidth]{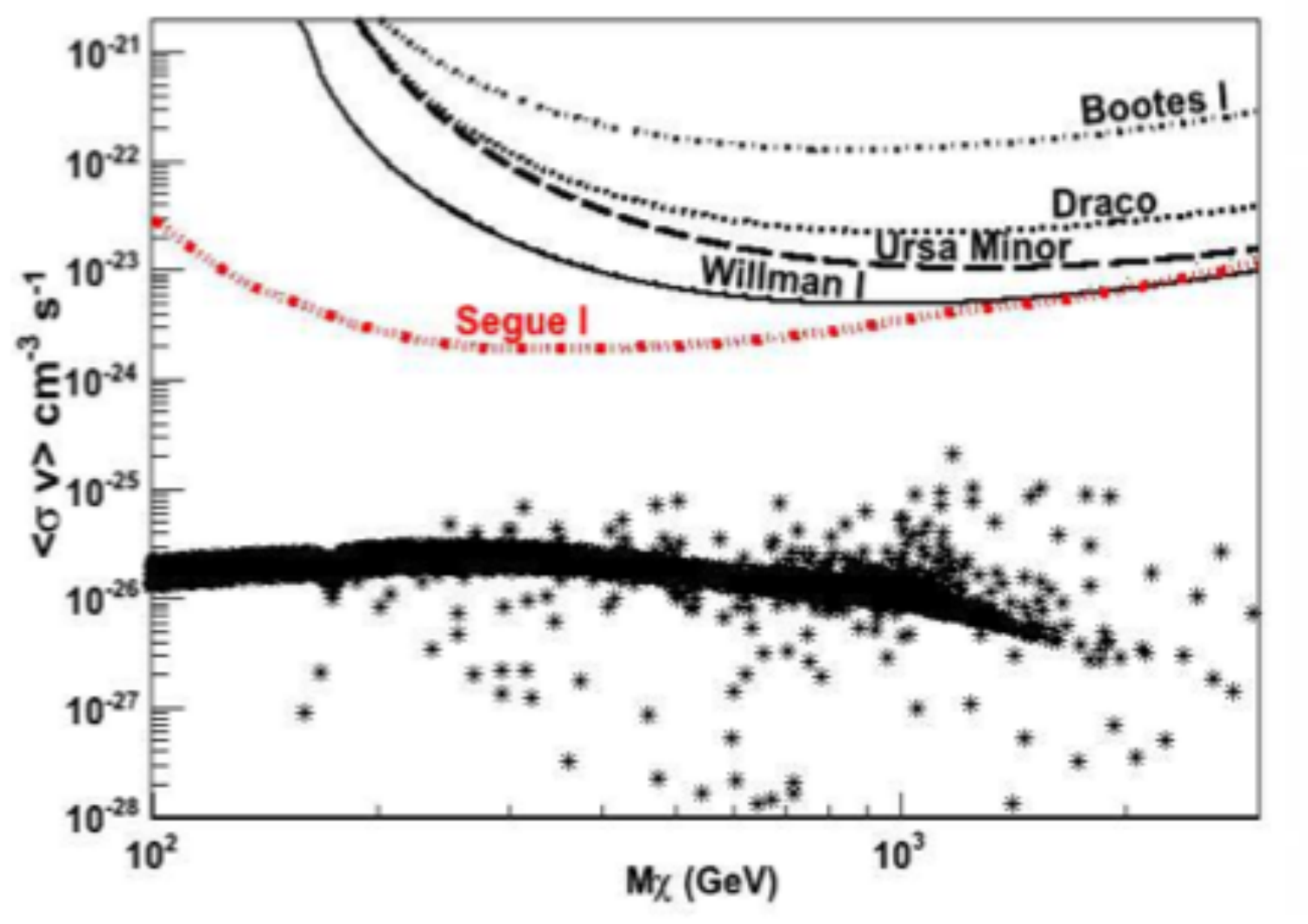}
	\caption{DM annilhation cross-section limits for $\sim$10-15 hours for four dSphs observations in black \cite{bib:VERITASCollab10} and for the $\sim$50 hrs observation of Segue 1 in red \cite{bib:VERITASCollab12}.  }
\end{figure}

\subsection{Event Weighting Method}

Improvements of IACT sensitivity can be obtained from a joint analysis of individual targets into a single limit and utilizing more of the individual event information.  An effort within the VERITAS collaboration do this is currently underway. This discussion only mentions neutralino annihilation, but could also be used to determine limits on the neutralino decay cross-section. The methodology of Geringer-Sameth and Koushiappas (2011) \cite{bib:GeringerSameth12} for joint \emph{Fermi}-LAT analysis of dSph data is being adapted to IACT data. In their analysis, every event in an region of interest (ROI) around each dwarf is assigned a weight, $w_{i}$, based on the event's energy, which dwarf the event comes from ($\nu$), and the angular separation from the center of the dwarf ($\theta$). The optimal form of the weights is derived from a likelihood ratio test. The likelihood test is designed to test the hypothesis of each event originating from DM annihilation ($H_{s+b}$). The null hypothesis ($H_{b}$) in this case is a of a pure cosmic-ray (CR) background without DM annihilation. The event weight takes the form:
\begin{equation}
w_{i}=\log{(1 + s_{i}/b_{i})}
\end{equation}

The expected number of background events, $b_{i}$ is determined from a spectrum of CR events for each data run. The expected number of DM annihilation events, $s_{i}$ detected from a dSph $\nu$ with an energy between $E$ and $E+dE$ and an angular separation in a solid angle interval $d\Omega(\theta)$ takes the form:

\begin{equation}
s_{i}=J_{\nu}\frac{<\sigma v>}{8\pi m^{2}_{\chi}}\frac{d N_{\gamma}(E) }{dE}t_{obs}A_{eff}(E) PSF(E,\theta) dE d\Omega(\theta).
\end{equation}

The weight plotted as a function of event energy and angular distance is shown in figure 2. The test statistic, $T$, a single number representing all the data used to test the DM hypothesis for all events, is the sum of $N$ weights:
\begin{equation}
T=\sum_{i=1}^{N}{w_{i}}=\sum_{i=1}^{N}\log{(1 + s_{i}/b_{i})}.
\end{equation} 

The point spread function, $PSF(E,\theta)$ (obtained from $\gamma$-ray simulations) is the probability per soild angle of detecting a photon of energy $E$ an angular distance $\theta$ away from the dwarf. 

A line search will be performed with this method, in which the annihilation spectrum is replaced by $2\delta(E-E_{\gamma})$ that is convolved with the VERITAS energy resolution ($\Delta E/E \sim20\%$). 

Given values of $m_{\chi}$ and $<\sigma v>$, a probability distribution function (PDF) is calculated which is used to test the hypothesis $H_{s+b}$. The PDF has the form of a compound Poisson distribution. The test statistic was chosen in a way such that larger values of $T$ indicate the presence of a signal. Confidence limits are found by testing several values of $m_{\chi}$ and $<\sigma v>$ and calculating $T$ at each step. If $T$ is larger than a critical value $T^{*}$ which represents the confidence level being tested (for example, 95\% confidence), then the hypothesis $H_{b}$ is rejected at that confidence. 


Figure 3 shows the expected limits over the lifetime of VERITAS using the weighting method. An one thousand hour exposure on Segue 1 and dSphs with similar $\it{J}$ factors are assumed, which could be an optimistic assumption. An energy cut of below 100 GeV and above 10 TeV was used. The effect of the VERITAS upgrade was simulated with by increasing the effective area for 750 hours of the total 1000 hours.  An additional factor of 2 to 3 could be obtained from advanced analysis techniques. 

\begin{figure}[t]
	\centering
	\includegraphics[width=0.5\textwidth]{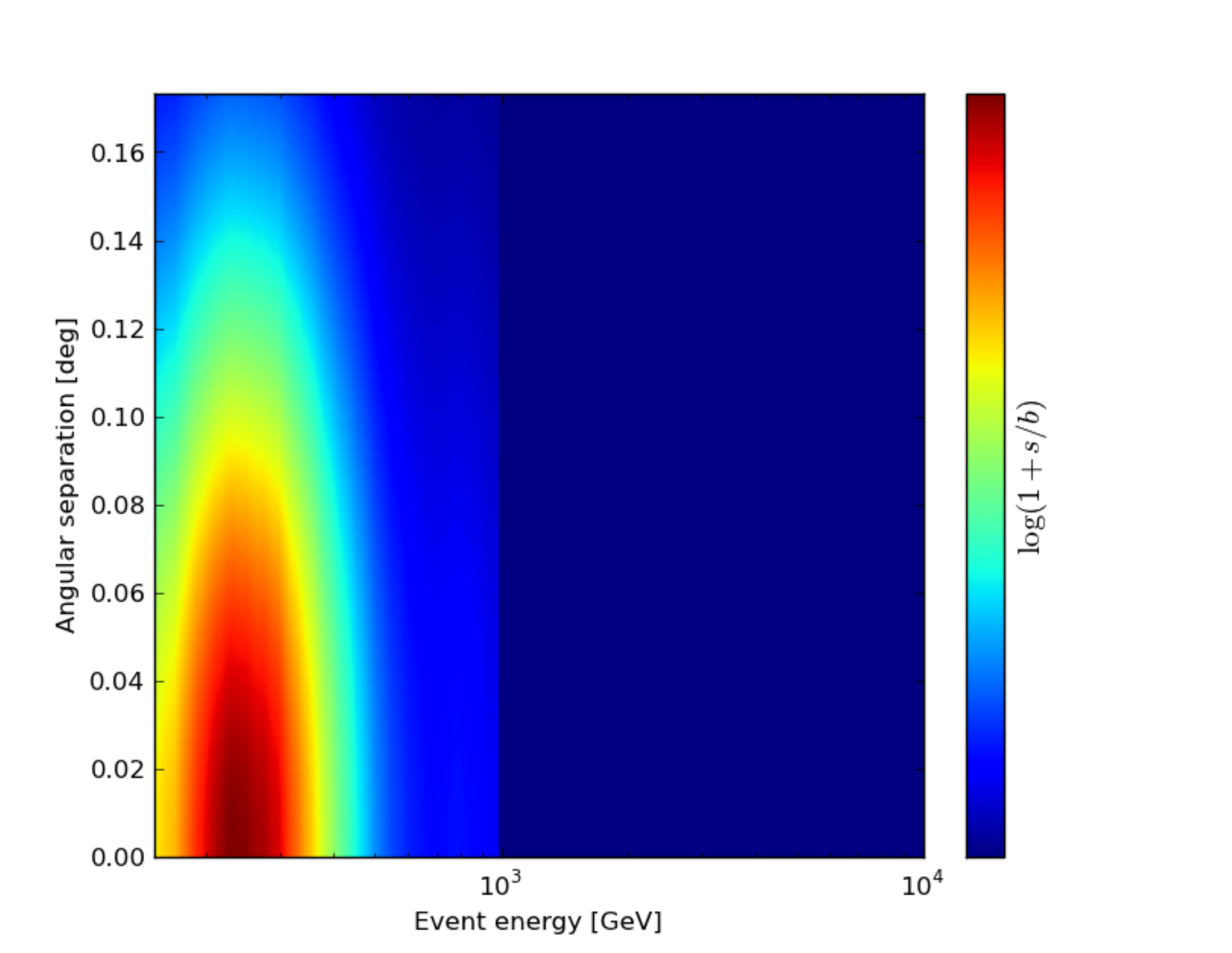}
	\caption{Plot showing the form of the event weight (Equation 2) as a function of event energy and angular distance to the dSph center. Units on the z-axis are arbitrary. The neutralino mass is 1 TeV.}
\end{figure}

This method, which was first used on \emph{Fermi}-LAT data, could easily be adapted for IACT data with the major exception of  selection of background (or OFF) region(s). Traditionally, IACT analysis selects background events (dominated by CRs), from either similarly sized regions around the tracking position which is offset from the source position (the reflected region method (RRM), or commonly known as the `wobble' analysis) or from an annulus around the ON region (the ring background method (RBM)). The RRM does not provide enough background events required to model the CR background spectrum for an accurate measurement of $b_{i}$, particularly for extended source analysis and/or sources with large exclusion regions caused by bright stars. The varying acceptance within the background regions for the RBM makes it unsuitable for spectral/flux calculations. Spectral calculations using the RBM would require energy-dependent acceptance functions, which would require binning in energy and increase overall systematic uncertainty\cite{bib:Berge08}. 

A third method, developed specifically for this analysis, selects background events from an annulus like the RBM, but the annulus is centered on the $\it{tracking}$ position (or center of the camera), as opposed to the $\it{source}$ position. An illustration of this method is shown in Figure 4. This gives roughly a factor of two greater useable background events, while the acceptance function within the chosen background region is relatively flat, since the acceptance is typically radially symmetric \cite{bib:Berge08}. The weight of each background event, $\alpha$, used to calculate $\gamma$-ray excess and significance \cite{bib:LiMa83} is determined by the ratio of the area of the ON region to the numerical integration of the area of each bin ( $\delta A_{off,i}$ ) of the background region of the sky map:

\begin{figure}[t]
	\centering
	\includegraphics[width=0.5\textwidth]{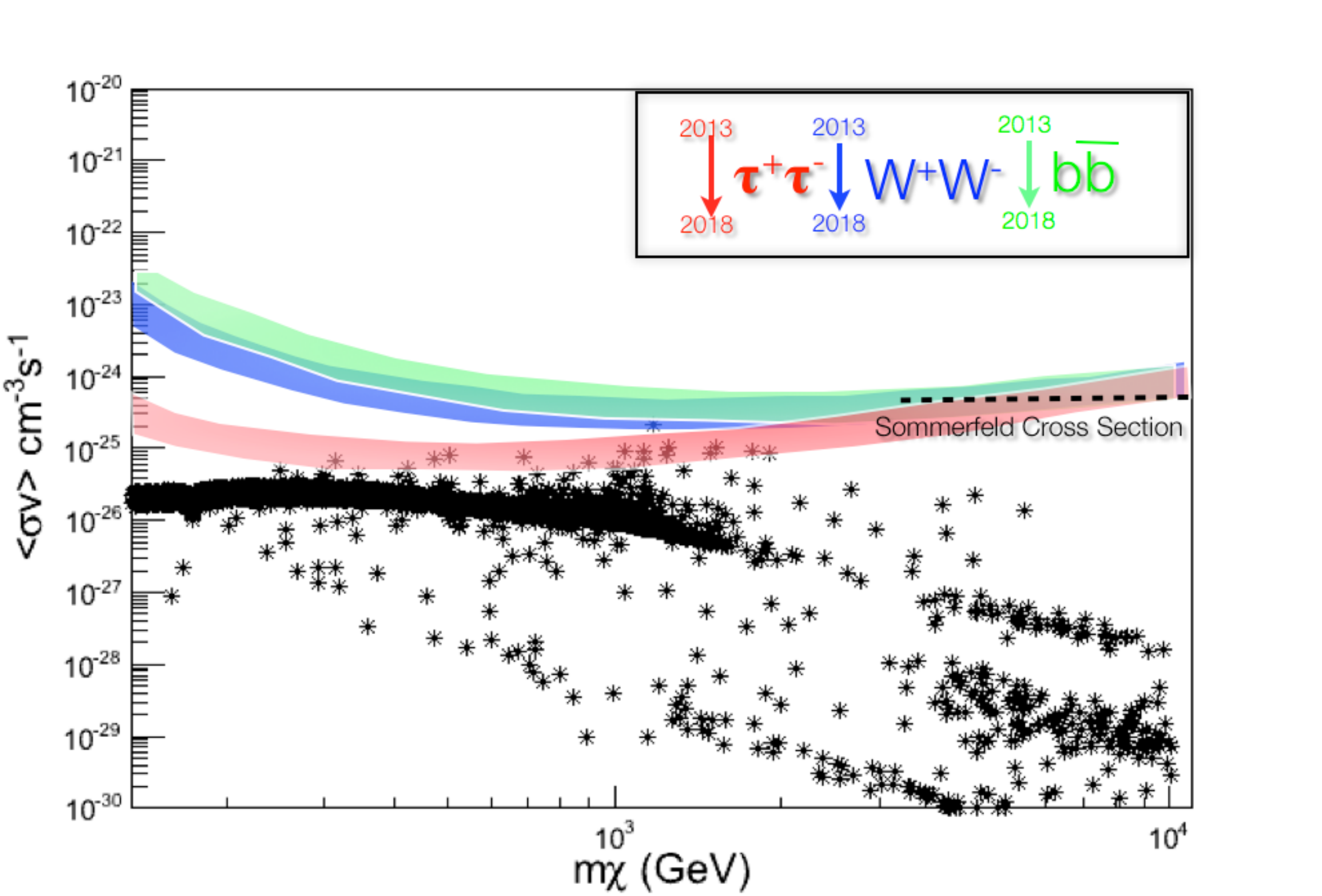}
	\caption{Expected limits from the VERITAS DM program to 2018 using the event weighting method. Each black dot represents a different derived cross section and mass from various models. Each black dot represents a different derived cross section and mass from various models. }
\end{figure}

\begin{equation}
\alpha= \frac{2\pi ( 1 - \cos ( \theta_{max} ) ) } {\sum{\delta A_{off,i} } },
\end{equation}

where  $\theta_{max}$ is the angular distance to the dSph location used to define the ON region. This method is nicknamed the `crescent' background method (CBM), since the ON region requires an exclusion region for the OFF region, making the OFF region shaped somewhat like a crescent. 
Tests of this method on the Crab Nebula and Segue 1 data has shown approximately the same background CR rates as the other background methods. Table 2 summarizes the comparison between the RRM and CBM methods. An independent calculation of $\alpha$ using MC verified the accuracy of equation 5 to within 1\%.
 
\begin{table*}[t]
	\centering
	
	\begin{tabular}[width=0.45\textwidth]{ | c || c | c || c | c | }
		\hline
		 & RRM (Crab) & CBM (Crab) & RRM (Segue) & CBM (Segue)\\
		\hline
		$N_{on}$  & 4749 & 4749 & 12361 & 12361\\ 
		\hline
		$N_{off}$ & 4646 & 7782 & 37137 & 93892 \\
		\hline
		$\bar{\alpha} $ & 0.20 & 0.12 & 0.35 & 0.13\\
		\hline
		$R_{bg} (CR/min)$ & 3.18$\pm$0.05 & 3.19$\pm$0.04& 2.55$\pm$0.13 & 2.51$\pm$0.08 \\
		\hline
		$R_{\gamma} (\gamma/min)$ & 13.39$\pm$0.24  & 13.35$\pm$0.24 & -0.12$\pm$0.03 &-0.03$\pm$0.02 \\
		\hline
		Significance & 79.8$\sigma$ & 84.1 $\sigma$ & -1.34$\sigma$ & -0.31 $\sigma$\\   
		\hline
	\end{tabular}
	\caption{ Summary of the test results comparing the RRM to the CBM. on 16 Crab Nebula (live time of 292.4 minutes) and 292 Segue 1 runs.}
\end{table*}

\begin{figure}[h]
	\centering
	\includegraphics[trim=10mm 70mm 10mm 10mm, clip=true, width=0.4\textwidth]{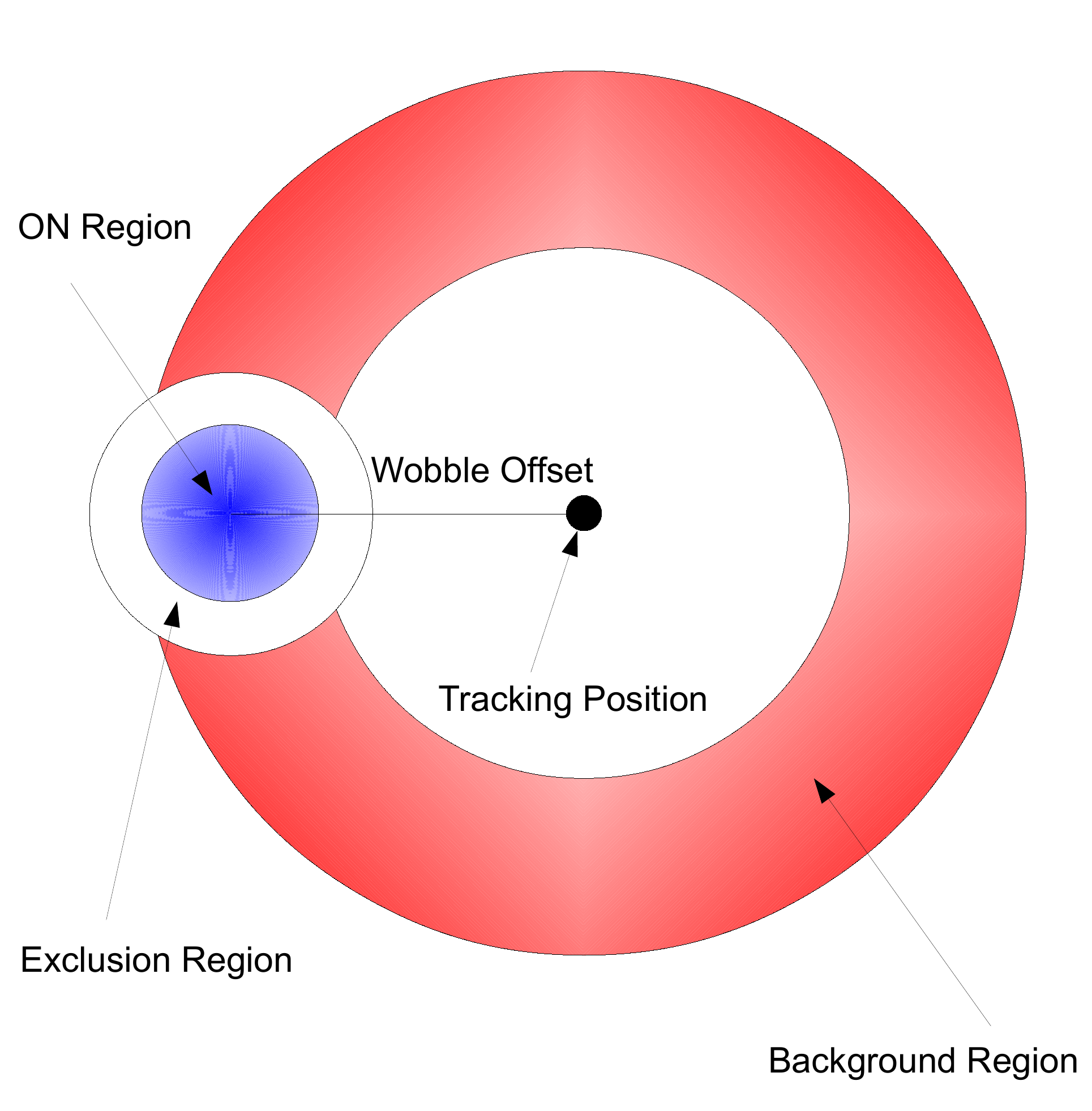}
	\caption{Example Illustration of the background method that will be used for the DM weighting analysis. The ON region is shaded in blue, while the OFF region is shaded red. Note that this figure is not drawn to any scale.}

\end{figure}

\section{Conclusions}

VERITAS is continuing to devote a significant portion of its observing schedule to DM targets, with dSphs as a key component of these observations. Roughly half of those observations have been devoted to Segue 1, which is the closest of the dSphs and has lowest obtainable energy threshold (see Table 1). The VERITAS upgrade aims to obtain a lower energy  threshold of $\sim$80 GeV, by employing faster FPGA-based pattern triggers \cite{bib:UpgL2} and new high-QE PMTs \cite{bib:UpgCam}, which should provide more low energy events, which should improve the DM limits at all neutralino masses. 

The weighting analysis represents a major future improvement in methodology for IACT DM analysis, which is still in the preliminary stages of implementation. Using the single event energy and position will provide a more accurate estimate of the cross-sections. For example, both the PSF and $J$ factor fall off as a function of distance from the dSph center, so events close to the center of the dSph position will be assigned greater weights. Events with low energy will be assigned greater weights than those of higher energy, since DM spectra cannot exceed the WIMP mass. Events with energies greater than the WIMP mass will be assigned weights of zero. The VERITAS camera upgrade will therefore improve limits at all WIMP masses, since it will provide more events at low energies. At the moment, the weighting method is being employed by dSphs observed with VERITAS, but it possible to combine not only different dSphs, but other DM targets, such as galaxy clusters or the Galactic Center with this method, and also different instruments, such as the \emph{Fermi} $\gamma$-ray telescope. Accuracy of the $J$ factor, which is the largest source of systematic uncertainty, will improve with more modeling of the DM density profiles and future spectroscopic surveys.

\vspace*{0.5cm}
\footnotesize{{\bf Acknowledgment:}{This research is supported by grants from the U.S. Department of Energy Office of Science, the U.S. National Science Foundation and the Smithsonian Institution, by NSERC in Canada, by Science Foundation Ireland (SFI 10/RFP/AST2748) and by STFC in the U.K. We acknowledge the excellent work of the technical support staff at the Fred Lawrence Whipple Observatory and at the collaborating institutions in the construction and operation of the instrument.}}

\end{document}